# A Novel Electrically Small Antenna Array Employing Opposite-Handed Chiral Parasitic Elements

Oleksandr Malyuskin[1],

[1] Centre for Wireless Innovation, Queen's University Belfast, Belfast, UK, o.malyuskin@qub.ac.uk

*Abstract*— **This paper presents a novel concept for electrically small antenna arrays (ESAAs) incorporating chiral parasitic elements of opposite handedness**. This configuration mitigates the detrimental effects of electromagnetic mutual coupling, which in conventional arrays causes a 180° phase shift between adjacent antenna currents when the element spacing is less than half a wavelength. The proposed approach is experimentally validated using a seven-element monopole ESAA with compact dimensions, specifically below half-wavelength in cross-section and one-sixth to one-fourth of a wavelength in vertical range. The antenna elements are spaced less than one-sixth wavelength apart, ensuring a highly compact footprint. Measurements show a –10 dB return-loss fractional bandwidth of 5–15% and a realised gain of 5–9 dBi, along with full 360° azimuthal beam-steering capability. The results confirm that employing oppositely handed chiral parasitic elements can significantly enhance performance in densely packed, electrically small antenna arrays.

## I. Introduction

Electrically small antennas (ESAs) are indispensable for modern wireless systems on miniaturised platforms. They play a key role in many technological areas, including Internet-of-Things (IoT) sensors, biomedical implants, compact radar units, and small satellite communications, where space, weight, and cost constraints are critical for the system performance. However, antenna miniaturisation inherently leads to reduced radiation efficiency and a narrow available bandwidth. The classical analyses of Wheeler, Chu, and Harrington established that as the antenna's electrical size decreases, its quality factor $Q$ increases sharply, limiting both bandwidth and efficiency [1]–[3]. The Bode–Fano theorem further constrains the performance of passive matching networks, imposing a fundamental tradeoff between bandwidth and reflection coefficient [4], [5].

A practical approach to overcome these inherent limitations is the use of electrically small antenna arrays (ESAAs)—multiple tightly spaced ESA elements configured so that the overall array size remains sub-wavelength. Although the individual elements are inefficient, the array can collectively achieve enhanced directivity, improved radiation resistance, and pattern control through constructive interference. With proper excitation control and mutual coupling management, ESAAs can exhibit superdirective behaviour, achieving directivities much greater than those of isolated ESAs of equivalent size [6], [7].

Recent studies have demonstrated that even two- and three-element ESAAs can achieve realised gains exceeding 7–10 dBi for $ka < 1$, where $k$ is a wavenumber and $a$ is a characteristic antenna spatial dimension, especially when Huygens-type (collocated electric and magnetic dipole) elements are employed [8], [9]. Such designs exploit strong inter-element coupling and precisely controlled excitation to synthesise current distributions that maximise the array's directivity without increasing its physical aperture. Modal and characteristic-mode analyses have further deepened understanding of the coupling mechanisms, enabling more systematic array synthesis, bandwidth optimisation, and reconfigurable operation [10], [11].

Despite these advances, ESAAs remain challenged by narrow operational bandwidth due to the inherently high-$Q$ nature of small elements and strong inter-element coupling. To address this, researchers have explored active and non-Foster matching circuits, which use negative-impedance converters (NICs) or time-varying reactive elements to neutralise the antenna's stored energy and broaden its impedance bandwidth [12]–[14]. Non-Foster networks break the classical Foster reactance constraint, providing reactances with negative slopes versus frequency. When integrated into small antennas or arrays, they can substantially expand the achievable bandwidth beyond passive Bode–Fano limits. While stability and noise remain practical design challenges, such active networks continue to show promise for next-generation, miniature broadband systems [12]–[14].

In parallel, metamaterial and metasurface technologies have emerged as powerful tools for enhancing the performance of electrically small antennas. Metamaterial shells and artificial magnetic conductor (AMC) substrates can confine and tailor near fields, improving impedance matching and radiation efficiency [15], [16]. Likewise, metasurface and high-index dielectric superstrates act as "metalenses," collimating radiation to improve realised gain and directivity without increasing size [17], [18]. Integrating these concepts with ESAAs enables novel compact arrays that achieve a balance among bandwidth, efficiency, and directivity, narrowing the gap between theoretical limits and practical realisations.

This paper presents, for the first time, a novel design methodology for electrically small antenna arrays that employs chiral parasitic elements of opposite handedness surrounding the active antennas. Traditionally, when the spacing between array elements is reduced below half a wavelength, strong mutual coupling between radiating elements causes a 180° phase shift, reducing radiation efficiency and substantially increasing radiation resistance.

In the proposed approach, introducing parasitic elements with opposite chirality induces electromagnetic interactions that reverse the magnetic field direction in neighbouring elements. This reversal minimises the phase shift between adjacent antenna currents, resulting in a significant improvement in both radiation efficiency and realised gain.

The concept is validated through performance data from a seven-element electrically small monopole array tested in both transmit and receive modes. In addition to improved return loss bandwidth, experimental results demonstrate conjugate (retrodirective) beamforming, highlighting the array's superdirective characteristics.

## II. ESAA Concept and Performance

### A. ESAA Geometry

The concept of mutual-coupling reduction using parasitic opposite-handedness chiral elements (helices) is explored in this paper using a 7-element monopole ESAA, Fig.1.

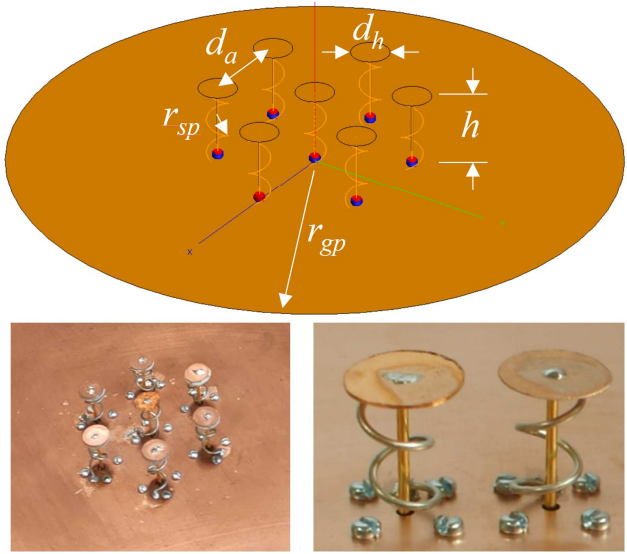

Fig.1. Antenna array geometry (top), fabricated ESAA (bottom left) and detailed view of adjacent antenna elements (bottom right).

In this design, a top-hat wire monopole array is fed by 50-Ohm coaxial sub-miniature version A (SMA) connector antenna ports. A spiral of opposite handedness surrounds each active antenna element, and the antenna elements are arranged in the face-centric hexagonal lattice (hexagonal close-packed lattice), which is optimal for dipole or monopole ESAA designs [19]. In the present ESAA design, monopole antenna height $h$ is in the range ($\lambda/8 - \lambda/4$), the top hat diameter is in the range ($\lambda/8 - \lambda/10$) and the antenna interelement spacings are in the range ($\lambda/8 - \lambda/4$). The ground plane diameter $r_{gp}$ is approximately $\lambda/2$, where is the radiation wavelength. The helix design offers flexibility between the spiral radius and the number of turns; the general rule is to keep the total length of the helix close to $\lambda/2$. The lower ends of the helices are connected to the ground plane.

### B. Return loss

A measured return loss of the 7-element ESAA is shown in Fig.2 for the ESAA designed for the S-band.

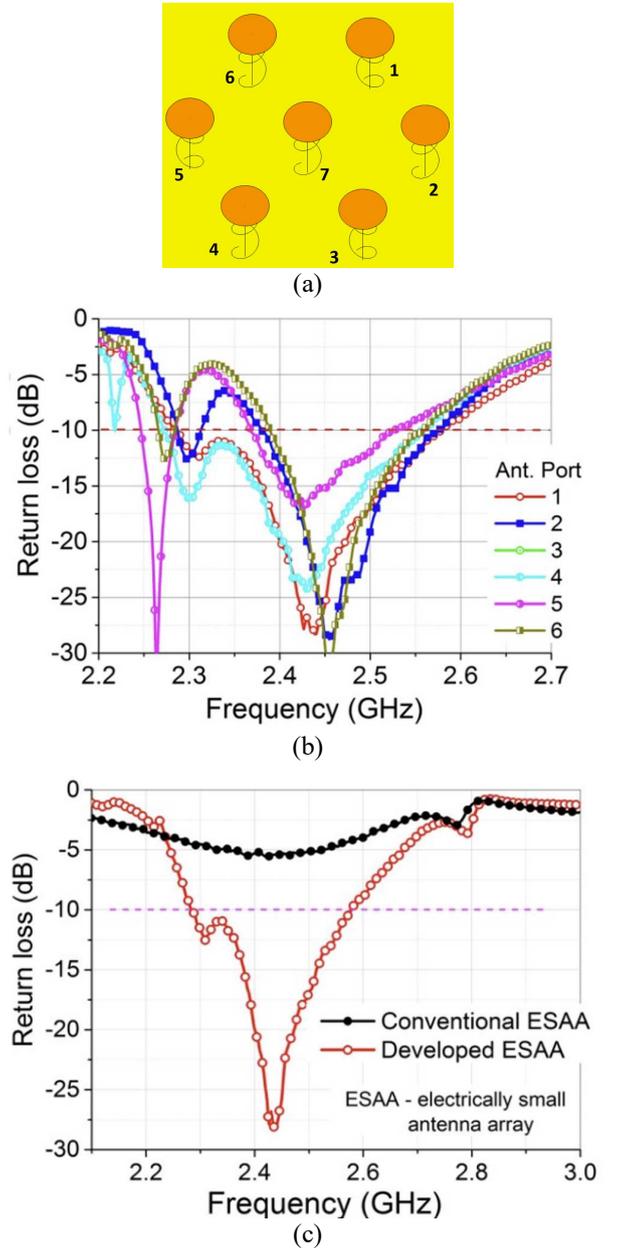

Fig.2. (a) ESAA ports numeration; (b) return loss of ESAA with parasitic helix elements (c) comparison between the developed ESAA and conventional ESAA composed of monopole antennas of the same size and geometry but without parasitic elements. The ESAA parameters: $h = \lambda/6$, $d_a = \lambda/6$, $r_{sp} = 2.54$mm, $d_h = \lambda/10$, $\lambda = 0.125$m.

The results in Fig. 2 show that the conventional monopole ESAA fails to achieve a return loss below -6dB, while the new ESAA ensures a return loss of -10dB in a wide fractional bandwidth of up to 15%. These results can be understood by examining the antenna current phase distribution, which reveals that the currents in the array remain well away from a 180-degree phase difference.

*C. Radiation Patterns*

The ESAA radiation patterns are generated using conjugate or retrodirective beamforming [20] and measured in the far-field range within the anechoic chamber.

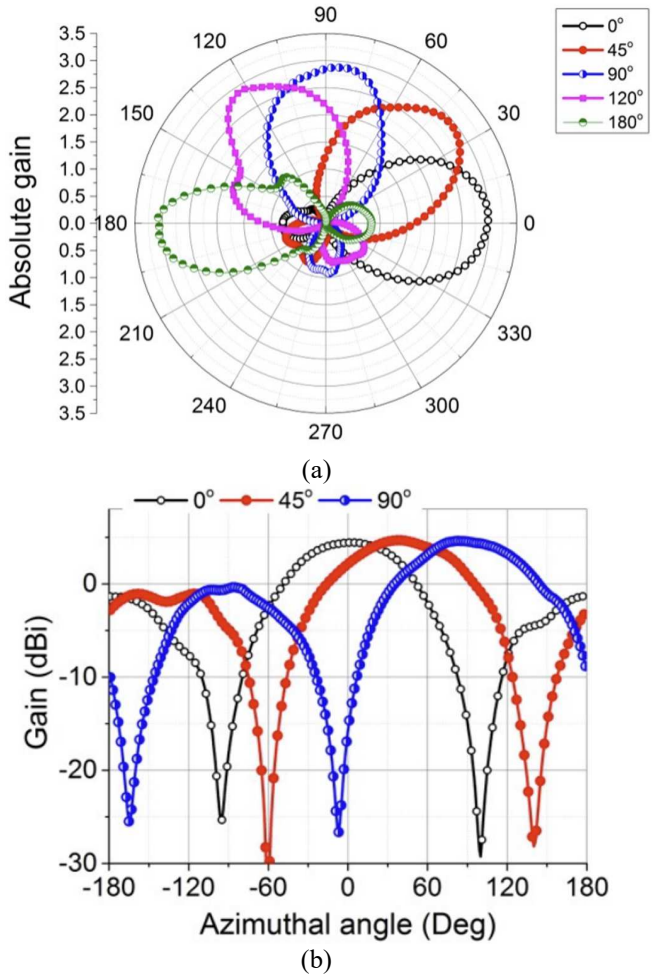

Fig.3. ESAA retrodirective radiation patterns in absolute (a) and logarithmic (b) scales. The inter-element antenna spacings are $\lambda/6$

In the conjugate beamforming (retrodirective) process, the receive voltages are measured for directions of arrival (DoA) of 0, 45, …, 180 degrees, phase-conjugated, and re-applied to the ESAA ports. The transmit retrodirective radiation patterns are shown in Fig. 3. It can be seen that the beam is accurately retrodirected to the original DoA with minimal beam squint. The data in Fig. 3(b) shows that the full beamwidth at half-maximum (FBHM) for the ESAA with $\lambda/3$ cross-section is less than 100 degrees, which suggests that the ESAA exhibits super-directive properties (in reference to $\lambda/2$ aperture generating FBHM of 125 degrees).

III. CONCLUSIONS

A new concept of an ESAA with parasitic chiral elements with opposite handedness has been proposed and explored experimentally. Experimental results show a significant improvement in the return loss performance, specifically, the 7-element monopole ESAA fractional bandwidth varies between 5% to 15 % for the antenna inter-element spacings in the interval of one-tenth to one-sixth of the radiation wavelength. It is demonstrated that retrodirective beamforming can be implemented using this type of ESAA, resulting in full azimuthal angular coverage with realised gain between 5-9dBi with minimal beam squint (less than 10 degrees) and superdirective radiation patterns with FBHM between 85 and 100 degrees. The proposed ESAA offer attractive capabilities in designing communication and electromagnetic sensing systems on ultracompact platforms.